\documentclass{cimento}
\usepackage{graphicx}
\usepackage{subfigure}
\usepackage{wrapfig}
\usepackage{caption}
\usepackage{mwe}
\usepackage{amsmath}

\def\Re{{\rm Re}} 
\def\Im{{\rm Im}}

\def\kv{{\bf{k}}}

\def\zu{\hat{\bf z}}

\def\kt{     { \bf{k}_{\rm T}  }    }

\def\kpt{  \kv'_{\rm T} }

\def\pt{     { \bf{p}_{\rm T}  }    }

\def\M{_{\rm{M}}}

\def\qbar{\bar{q}}

\def\gl{G_{\rm L}}
\def\gt{G_{\rm T}}

\def\sigmaT{\boldsymbol{\sigma}_{\rm T}}
\def\kpttilde{\tilde{\textbf{k}}'_{\rm T}}

\def\VT{\textbf{V}_{\rm T}}
\def\VL{V_{\rm L}}
\def\ktbar{\bar{\textbf{k}}_{\rm T}}
\def\kptbar{\bar{\textbf{k}}'_{\rm T}}
\def\Pt{\textbf{P}_{\rm T}}
\def\kptbartilde{\tilde{\bar{\textbf{k}}}'_{\rm T}}
\def\zq{\hat{\textbf{z}}_q}
\def\zqbar{\hat{\textbf{z}}_{\qbar}}

\def\helQuark{\lambda_q}
\def\helAntiQuark{\lambda_{\qbar}}
\def\M{\hat{\mathcal{M}}}

\def\Iq{1^q}

\def\sigmaXq{\sigma_x^q}
\def\sigmaYq{\sigma_y^q}
\def\sigmaZq{\sigma_z^q}
\def\Iqbar{1^{\qbar}}

\def\sigmaXqbar{\sigma_x^{\qbar}}
\def\sigmaYqbar{\sigma_y^{\qbar}}
\def\sigmaZqbar{\sigma_z^{\qbar}}

\def\Trqq{\rm{Tr}_{q\qbar}}

\title{Modeling spin effects in electron-positron annihilation to hadrons}
\author{A.~Kerbizi\from{ins:a}\thanks{albi.kerbizi@ts.infn.it},
X.~Artru\from{ins:b}
}
\instlist{\inst{ins:a} Dipartimento di Fisica, Universit\`a  degli Studi di Trieste and INFN Sezione di Trieste
 \inst{ins:b} Universit\'e de Lyon, Institut de Physique des deux Infinis (IP2I Lyon), Universit\'e Lyon 1 and CNRS}

\begin{document}

\maketitle

\begin{abstract}
The string+${}^3P_0$ model of spin-dependent hadronization is applied to the fragmentation of a string stretched between a quark and an antiquark with entangled spin states, assumed to be produced in the $e^+e^-$ annihilation process. The model accounts systematically for the spin correlations in the hadronization chain and is formulated as a recursive recipe suitable for the implementation in a Monte Carlo event generator. The recipe is applied to the production of two back-to-back pseudoscalar mesons produced in $e^+e^-$ annihilation, and it is shown to reproduce the form of the azimuthal distribution of the hadrons as expected in QCD.
\end{abstract}

\section{Introduction}
The $e^+e^-$ annihilation to hadrons is an important process for hadronization studies. In particular it allows to access the still poorly known spin-dependent fragmentation functions (FFs), such as the Collins FF \cite{collins}. The latter can be accessed by looking at the amplitude of a particular modulation in the azimuthal distribution of two back-to-back hadrons, known as the Collins asymmetry \cite{Boer:2008-e+e-}. Also, the combined analysis of $e^+e^-$ data and data from semi-inclusive deep inelastic scattering (SIDIS) allows to obtain information on both the FFs and on the partonic structure of the nucleons.

In this article, we use the string+${}^3P_0$ model of hadronization \cite{kerbizi-2018,Kerbizi:2021M20}, briefly recalled in Sect. \ref{sec:string+3P0}, to describe the quark spin effects in $e^+e^-$ annihilation at leading order. The string+${}^3P_0$ model was successfully applied to the description of transverse spin effects in SIDIS, either via standalone simulations \cite{kerbizi-2018,Kerbizi:2021M20} or with the implementation in the Pythia generator \cite{Kerbizi:2021StringSpinner,Kerbizi:2023cde}. The description of $e^+e^-$ annihilation is, however, more intricate than SIDIS due to the correlated spin states of the quark $q$ and the antiquark $\qbar$ stretching the string to be fragmented. The application of the string+${}^3P_0$ model to $e^+e^-$ is described in Sect. \ref{sec:recipe}. The $q\qbar$ pair is assumed to be produced by the process $e^+e^-\rightarrow q\qbar$ via the exchange of a virtual photon. The spin-correlations are propagated along the fragmentation chain by using the splitting amplitude of the string+${}^3P_0$ model, and by employing the Collins-Knowles (CK) recipe \cite{collins-corr,knowles-corr} to take into account the spin-entanglement at each step of the fragmentation chain. The resulting model is formulated as a recursive recipe suitable for the implementation in MC event generators, which is presented in Sect. \ref{sec:recipe}. In Sect. \ref{sec:application} we apply the recipe to the production of two back-to-back hadrons in $e^+e^-$. 
Finally, the conclusions are given in Sect. \ref{sec:conclusions}.

\section{The string+${}^3P_0$ model}\label{sec:string+3P0}
In the string+${}^3P_0$ model, the hadronization process $q\qbar\rightarrow h_1,h_2,\dots$ is the fragmentation of a straight string stretched between $q$ and $\qbar$, here assumed to have momenta along the $\zq\equiv \zu$ axis and the $\zqbar\equiv-\zu$ axis, respectively. The string fragmentation can be viewed as the recursive iteration of elementary splittings $q\rightarrow h + q'$, with $q$, $h$ and $q'$ being the fragmenting quark, the leftover quark and the emitted hadron, respectively. $h$ can be either a pseudoscalar (PS) or a vector meson (VM). The splitting is described in flavour, momentum and spin space by the splitting amplitude $T_{q',h,q}(Z_+,\pt;\kt)$. $k$, $p$ and $k'$ indicate the four momenta of $q$, $h$ and $q'$, respectively. The transverse components with respect to the string axis are labelled by $\kt$, $\pt=\kt-\kpt$ and $\kpt$. For-momentum conservation implies $p=k-k'$. The longitudinal splitting variable is defined as $Z_+=p^+/k^+$, $v^{\pm}=v^0\pm v^z$ being the lightcone components of a generic four-vector $v$. The expression for the splitting amplitude in the string+${}^3P_0$ model is \cite{Kerbizi:2021M20}
\begin{eqnarray}\label{eq:T}
T_{q',h,q}(M_h^2,Z_+,\pt;\kt) = \left[F^{\rm Lund}_{q',h,q}(M_h^2,Z_+,\pt;\kt)\right]^{1/2} \times \left[ \mu + i\sigmaT\cdot\kpttilde\right] \times \Gamma_{h,s_h}.
\end{eqnarray}
The function $F^{\rm Lund}_{q',h,q}$ describes the elementary splitting in flavour and momentum space (including the squared mass $M_h^2$, if $h$ is a resonance) \cite{Kerbizi:2021M20}, and it indicates the spin-less splitting function of the Lund Model \cite{Lund1983}. The $2\times 2$ matrix $\Delta_{q'}(\kpt)\equiv\mu + i\sigmaT\cdot\kpttilde$ implements the ${}^3P_0$ mechanism of quark pair production at string breaking. It depends on the complex parameter $\mu=\Re(\mu) + i\Im(\mu$) and on the transverse vector $\kpttilde=\zq\times\kpt$. $\Im(\mu)$ ($\Re(\mu))$ is responsible for transverse (longitudinal) spin effects, and has dimensions of mass. The $2\times 2$ matrix $\Gamma_{h,s_h}$ implements the coupling of $q$ and $q'$ with $h$. If $h$ is a PS meson it is $\Gamma_{h}=\sigma_z$. If $h$ is a VM it is $\Gamma_{h,\textbf{V}}=\gt\sigmaT\sigma_z\cdot\VT^*+1\,\gl\,\VL^*$, where $\textbf{V}=(\VT,\VL)$ is the polarization vector of the VM. $\gt$ and $\gl$ are two complex constants, which couple the quarks to a VM with transverse and longitudinal polarization with respect to the string axis, respectively.

The hadronization $q\qbar\rightarrow h_1,h_2,\dots$ can be viewed equivalently as iterations of elementary splittings $\qbar\rightarrow H +\qbar'$ of antiquarks, owing to the left-right symmetry \cite{Lund1983}. The momenta of $\qbar$, $H$ and $\qbar'$ are indicated by $\bar{k}$, $P$ and $\bar{k}'$, respectively. Their transverse components are indicated by $\ktbar$, $\Pt$ and $\kptbar$, respectively. Also, we define $Z_-=P^-/\bar{k}^-$. Four-momentum conservation gives $P=\bar{k}-\bar{k}'$. The splitting amplitude $T_{\qbar',H,\qbar}(Z_-,\Pt;\ktbar)$ associated to the splitting $\qbar\rightarrow H+\qbar'$ can be obtained from Eq. (\ref{eq:T}) with the substitutions $\lbrace q,h,q'\rbrace \rightarrow \lbrace\qbar,H,\qbar'\rbrace$, $Z_+\rightarrow Z_-$ and $\lbrace \kt,\pt,\kpt\rbrace\rightarrow \lbrace \ktbar, \Pt, \kptbar\rbrace$. In particular, it is $\Delta_{\qbar'}(\kptbar)\equiv\mu+i\sigmaT\cdot\kptbartilde$, where we have defined $\kptbartilde=\zqbar\times \kptbar$. 

\section{Spin-dependent string fragmentation in $e^+e^-$ annihilation}\label{sec:recipe}
\subsection{Physical ingredients}\label{sec:ingredients}
\begin{figure}[tbh]
\centering
\begin{minipage}[b]{0.49\textwidth}
\hspace{-0.8em}
\includegraphics[width=1.0\textwidth]{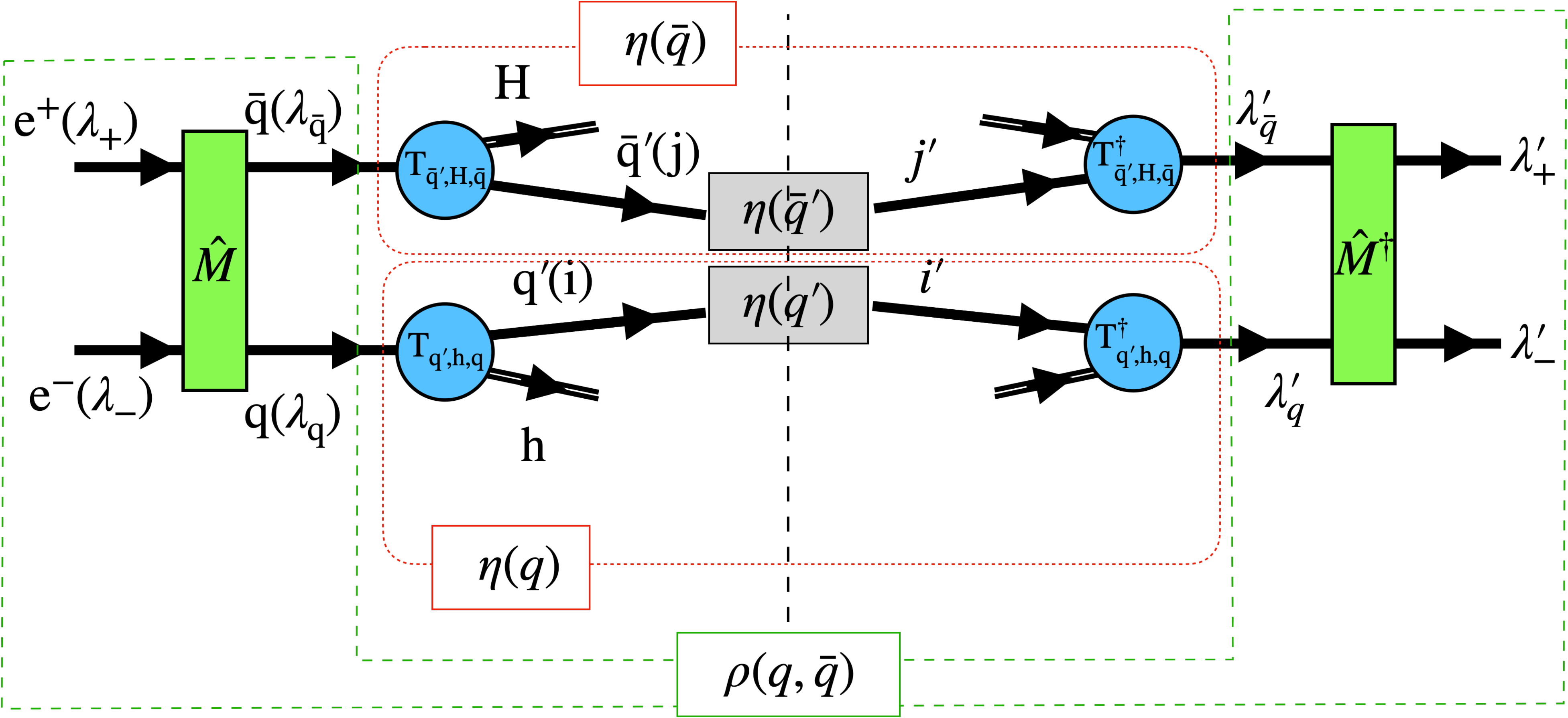}
\end{minipage}
\begin{minipage}[b]{0.49\textwidth}
\includegraphics[width=1.0\textwidth]{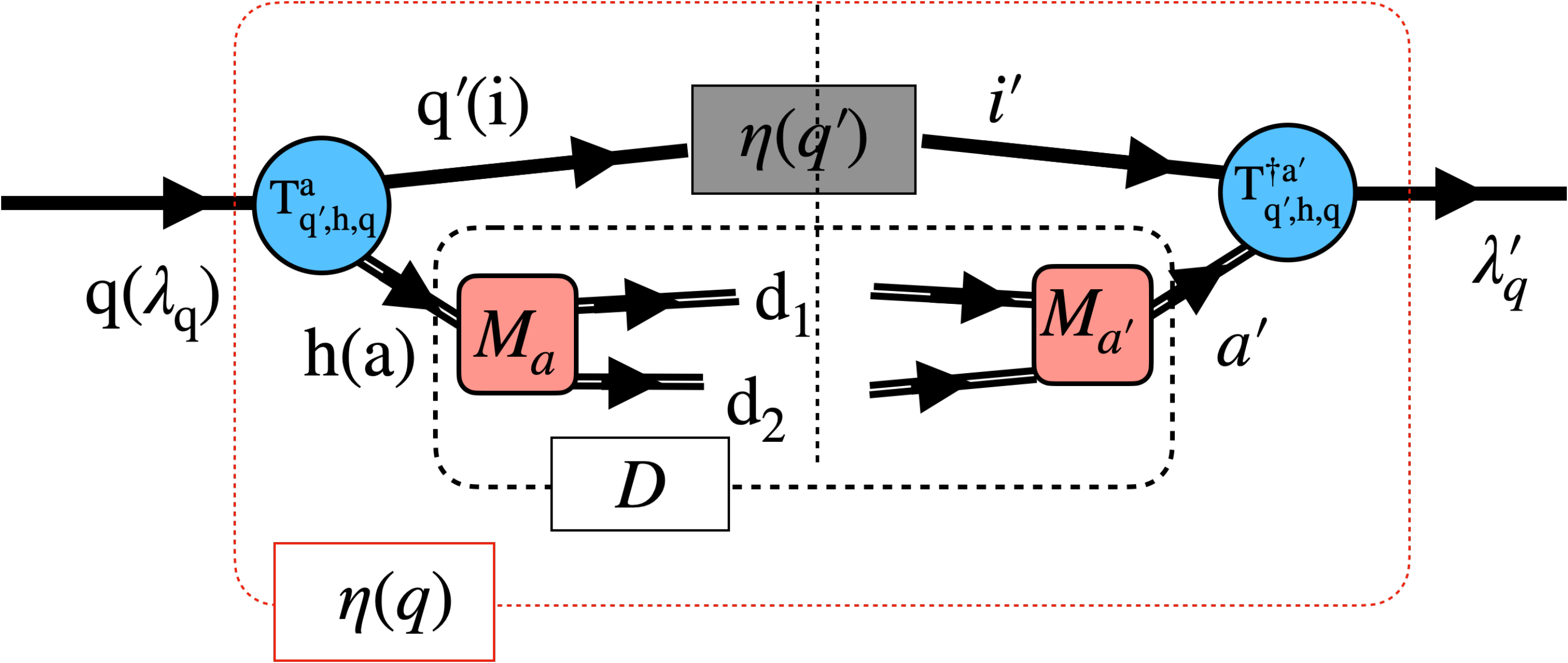}
\end{minipage}
\caption{Left: Unitarity diagram for $e^+e^-\rightarrow q\qbar\rightarrow h\,H\,X$. Right: Emission of a VM from the quark.}
\label{fig:unitary diagram}
\end{figure}

Following the factorization theorem \cite{Collins:1981uk}, we divide the process $e^+e^-\rightarrow h\,H\,X$ into the hard sub-process $e^+e^-\rightarrow q\qbar$, where the quark pair $q\qbar$ is created, and in the soft process $q\qbar\rightarrow h_1\, h_2\, ..$, where the quark pair hadronizes into the final state hadrons $h_1\, h_2\, ..$. 
The $e^+$ and $e^-$ are taken unpolarized. In Fig. \ref{fig:unitary diagram} (left) is shown the unitary diagram associated to the annihilation $e^+e^-\rightarrow h\,H\,X$, where the hadrons $h$ and $H$ are emitted by the splittings $q\rightarrow h + q'$ and $\qbar\rightarrow H+\qbar'$, respectively. To the unitary diagram we associate the sum of squared amplitudes
\begin{equation}\label{eq:A^2 e+e- simplified}
\begin{aligned}
   &\mathcal{I}= \langle |\M|^2\rangle \times \langle i|T_{q',h,q}|\helQuark\rangle \, \langle j |T_{\qbar',H,\qbar}|\helAntiQuark\rangle\, \langle\helQuark,\helAntiQuark|\rho(q,\qbar)|\helQuark',\helAntiQuark'\rangle\, \langle \helQuark'|T^{\dagger}_{q',h,q}|i'\rangle\,\langle \helAntiQuark'|T^{\dagger}_{\qbar',H,\qbar}|j'\rangle\\
   &\times \langle i'|\eta(q')|i\rangle\,\langle j'|\eta(\qbar')|j\rangle \equiv \langle |\M|^2\rangle\,\Trqq\left[\rho(q,\qbar)\,\eta(q)\otimes \eta(\qbar)\right],
\end{aligned}
\end{equation}
where we have indicated by $\helQuark$ ($\helAntiQuark$) the helicity index of $q$ ($\qbar$), and by $i$ ($j$) the spin index of $q'$ ($\qbar'$). The indices in the complex conjugated amplitude include in addition a prime symbol, as shown in Fig. \ref{fig:unitary diagram}. $\langle |\M|^2\rangle$ is the spin-averaged squared amplitude associated to the hard scattering $e^+e^-\rightarrow q\qbar$.

In the first line in Eq. (\ref{eq:A^2 e+e- simplified}), we introduced the joint spin density matrix $\rho(q,\qbar)$ of the $q\qbar$ pair, which implements the initial spin correlations between the quarks. It reads $\rho(q,\qbar)=\left[\Iq\otimes \Iqbar - \sigmaZq\otimes \sigmaZqbar + \hat{a}_{\rm NN}(\theta)\,(\sigmaXq\otimes \sigmaXqbar + \sigmaYq\otimes \sigmaYqbar)\right]/4$, where $\sigma_{\nu}^{Q}$ indicates the Pauli matrix along the axis $\nu=0,x,y,z$ ($0$ referring to the identity matrix) in the helicity frame of $Q=q,\qbar$. The quark helicity frames are defined by the axes $\hat{\textbf{z}}_Q\propto\textbf{k}_Q$, $\hat{\textbf{y}}_Q\propto\textbf{p}_-\times\hat{\textbf{z}}_Q$ and $\hat{\textbf{x}}_Q\propto\hat{\textbf{y}}_Q\times\hat{\textbf{z}}_Q$, with $Q=q,\qbar$, $\textbf{p}_-$ being the $e^-$ momentum and $\textbf{k}_Q$ the momentum of $Q$. The (anti)quark helicity frame will be indicated shortly by (A)QHF. The quantity $\hat{a}_{\rm NN}(\theta)=\sin^2\theta/(1+\cos^2\theta)$ induces transverse-spin correlations among the quarks, with $\theta$ being the angle between $\textbf{p}_-$ and $\textbf{k}_q$.

The $2\times 2$ matrix $\eta$ carries the spin information coming "backwards in time" form the analysis of particles further produced in the fragmentation chain, and it is referred to as the "acceptance matrix". If the particles are not analyzed, it is $\eta=1$. The expressions for $\eta(q)$ and $\eta(\qbar)$, obtained by comparing the top and bottom lines in Eq. (\ref{eq:A^2 e+e- simplified}), are $\eta(q)=T^{\dagger}_{q',h,q}\,\eta(q')\,T_{q',h,q}$ and $\eta(\qbar)=T^{\dagger}_{\qbar',H,\qbar}\,\eta(\qbar')\,T_{\qbar',H,\qbar}$, respectively, for PS meson emission. For VM emission, e.g., from $q$, the quark line in the left diagram in Fig. \ref{fig:unitary diagram} is replaced by the diagram on the right-hand side of the figure. $\eta(q)$ becomes then $\eta(q)=T^{a',\dagger}_{q',h,q}\,\eta(q')\,T^{a}_{q',h,q}\,D_{a'a}$, where $a$ ($a'$) indicates the spin index of the VM in the amplitude (complex conjugated amplitude). $T^a_{q',h,q}$ is obtained writing the splitting amplitude in Eq. (\ref{eq:T}) as $T_{q',h,q}=T^a_{q',h,q}\,\textbf{V}_a$. $D_{a'a}$ is the decay matrix of the VM \cite{Kerbizi:2021M20}. An analogue expression holds in the case of a VM emission from $\qbar$.

\subsection{The recursive recipe}
First, the hard scattering $e^+e^-\rightarrow q\qbar$ must be set up by using the differential cross section $d\hat{\sigma}(q\qbar)/d\cos\theta\propto \langle |\M|^2\rangle$. The latter is used to draw the flavours of the quarks, and the four-momenta of the quarks, $e^+$ and $e^-$. To explicitly write the momenta, and for the following steps, let us use as the reference frame the QHF. Then the string stretched between $q$ and $\qbar$ can be fragmented by the steps:
\begin{itemize}\setlength\itemsep{0em}
    \item[1.] Select with equal probability if the splitting is to be taken from the $q$ or $\qbar$ side.
    \item[2.] \textbf{For a splitting from the $q$ side}, assume no information is coming from the $\qbar$ end, thus $\eta(\qbar)=1_{\qbar}$ and $\eta(q')=1_{q'}$. If $h=\rm PS$, generate $h$ using the probability density (splitting function) $dP_{q\rightarrow h+q'}/dZ_+Z_+^{-1}d^2\pt=F_{q',h,q}=\rm{Tr}_{q'\qbar}[\textbf{T}_{q',h,q}\,\rho(q,\qbar)\,\textbf{T}^{\dagger}_{q',h,q}]$, obtained inserting the expression of $\eta(q)$ (see Sect. \ref{sec:ingredients}) in Eq. (\ref{eq:A^2 e+e- simplified}). Also, we defined $\textbf{T}_{q',h,q}=T_{q',h,q}\otimes 1_{\qbar}$.
    
    If $h=\rm VM$, apply the following additional steps. (a) Calculate the spin density matrix of $h$ using $\rho_{aa'}\propto \rm{Tr}_{q'\qbar}[\textbf{T}^a_{q',h,q}\,\rho(q,\qbar)\,\textbf{T}^{a'\,\dagger}_{q',h,q}]$, and use it to perform the polarized decay in the rest frame of the VM following Ref. \cite{Kerbizi:2021M20}. Afterwards boost the decay products in the QHF. (b) Following the CK recipe, evaluate the decay matrix $D_{a'a}$ of the VM (see Ref. \cite{Kerbizi:2021M20} for the explicit expressions).
    
    \textbf{For a splitting from the $\qbar$ side}, assume no information coming from the $q$ side, thus $\eta(q)=1_q$ and $\eta(\qbar')=1_{\qbar'}$. For $H=\rm PS$, generate $H$ using the probability density $dP_{\qbar\rightarrow H+\qbar'}/dZ_-Z_-^{-1}d^2\Pt=F_{\qbar',H,\qbar}=\rm{Tr}_{q\qbar'}[\textbf{T}_{\qbar',H,\qbar}\,\rho(q,\qbar)\,\textbf{T}^{\dagger}_{\qbar',H,\qbar}]$. This expression is obtained inserting the expression for $\eta(\qbar)$ (see Sect. \ref{sec:ingredients}) in Eq. (\ref{eq:A^2 e+e- simplified}). Also, we defined $\textbf{T}_{\qbar',H,\qbar}=1_q\otimes T_{\qbar',H,\qbar}$.
    For $H=\rm VM$, the analogue steps of (a)-(c) apply. Express all hadron momenta in the QHF.
    
    If the exit condition, which can be taken the same as in the spin-less Lund Model \cite{Lund1983}, is satisfied, go to step 4. Otherwise go to step 3.
    
    \item[3.] If the splitting was taken from the $q$ side, evaluate the spin density matrix $\rho(q',\qbar)$ using the expression $\rho(q',\qbar)\propto \textbf{T}_{q',h,q}\,\rho(q,\qbar)\,\textbf{T}^{\dagger}_{q',h,q}$ for the case $h=\rm PS$. This is obtained inserting $\eta(q)$ (see Sect. \ref{sec:ingredients}) in Eq. (\ref{eq:A^2 e+e- simplified}) without taking the trace. Using the CK recipe, for $h=\rm VM$ the expression is $\rho(q',\qbar)\propto \textbf{T}^{a}_{q',h,q}\,\rho(q,\qbar)\,\textbf{T}^{a' \dagger}_{q',h,q}\,D_{a'a}$. If the splitting was taken from the $\qbar$ side, $\rho(q,\qbar')$ can be obtained in an analogue way using $\textbf{T}_{\qbar',H,\qbar}$ instead of $\textbf{T}_{q,h,q'}$. Return to step 1.

    \item[4.] Assuming the last quark pair to be hadronized is $q_l\qbar_L$, emit the last two hadrons $h$ and $H$ by one further splitting. With equal probability, either generate $h$ from the $q_l$ side by using $F_{q',h,q_l}$ (see step 2) and construct $H=\qbar_L\qbar'$ using momentum conservation, or generate $H$ from the $\qbar_L$ side using $F_{\qbar',H,\qbar_L}$ (see step 2) and construct $h=q_l\qbar'$ using momentum conservation.
\end{itemize}

Steps 1-3 are iterated recursively until the condition for the termination of the fragmentation chain is called. At each step the spin correlations are propagated using quantum mechanical rules. This allows to account for correlations between the momenta of hadrons emitted in the $q$ and $\qbar$ jets. Being recursive, the proposed recipe is suitable for the implementation in a MC event generator. Also, since the recipe does not depend on the production mechanism of the $q\qbar$ pair, it can used for other processes as well.

\section{Application of the model to $e^+e^-\rightarrow h\,H\,X$}\label{sec:application}
As a proof-of-principle calculation, we evaluate the probability distribution for the process $e^+e^-\rightarrow h\, H\, X$, where $h$ is emitted in the splitting $q\rightarrow h+q'$ and $H$ is emitted in the splitting $\qbar\rightarrow H+\qbar'$. The hadrons are thus produced back-to-back in the $e^+e^-$ center of mass frame, each of them being associated with a different quark jet. According to the recipe in Sect. \ref{sec:recipe}, the probability associated to this process is obtained as the product of the probabilities for i) creating the $q\qbar$ pair, ii) emitting $h$ from $q$, and iii) emitting $H$ from $\qbar$ given the emission at ii). Thus, it is $dP_{e^+e^-\rightarrow h\,H\,X} \propto d\hat{\sigma}(q\qbar)\times F_{q',h,q}\times F_{\qbar',H,\qbar}$. Using the expression for the splitting amplitude in Eq. (\ref{eq:T}), and assuming $h=\rm PS$ and $H=\rm PS$, we obtain
\begin{equation}
\frac{dP_{e^+e^-\rightarrow h\,H\,X}}{d\cos\theta\,d\phi_h\,d\phi_H\,..} \propto (1+\cos\theta^2)\,\left[1+\hat{a}_{\rm NN}(\theta)\,\frac{2\Im(\mu)\,p_{\rm T}}{|\mu|^2+p^2_{\rm T}}\,\frac{2\Im(\mu)\,P_{\rm T}}{|\mu|^2+P^2_{\rm T}}\,\cos(\phi_h+\phi_H)\right],
\end{equation}
where the dots in the denominator indicate the differentials $dZ_{+}Z_+^{-1}\,dZ_-Z_-^{-1}\,dp^2_{\rm T}\,dP^2_{\rm T}$. As can be seen, the string+${}^3P_0$ model gives a $\cos(\phi_h+\phi_H)$ modulation in the distribution of the azimuthal angles $\phi_h$ of $h$ and $\phi_H$ of $H$, expressed in the QHF. This modulation is expected from QCD \cite{Boer:2008-e+e-}. The amplitude has been measured by different experiments and it is referred to as the Collins asymmetry \cite{belle-spin-asymmetries,babar}. Note that the amplitude predicted by the string+${}^3P_0$ model has positive sign, as in the data. In order to obtain the full Collins asymmetry for back-to-back hadrons in simulations of $e^+e^-$ annihilation, however, the MC implementation of the model is needed.

\section{Conclusions}\label{sec:conclusions}
We have applied the string+${}^3P_0$ model of spin-dependent hadronization to the description of $e^+e^-$ annihilation to hadrons. The new model, built at the amplitude level to preserve quantum mechanical properties such as positivity and entanglement, is formulated as a recursive recipe suitable for the implementation in a MC event generator. A proof-of-principle calculation has been performed by applying the model to the production of two back-to-back hadrons in $e^+e^-$ annihilation. The calculation shows that the model reproduces the Collins asymmetry in the distribution of the final state hadrons expected in QCD, and that it gives the same sign of the asymmetry as in the data. For more quantitative simulation results, the implementation of the model in a MC generator is needed. The 
work on the MC implementation is ongoing and will be presented in a future article.




\bibliographystyle{varenna}
\bibliography{main.bib}

\end{document}